# Exceptional magnetic anomalies in TbIrIn$_5$

Ayusa Aparupa Biswal[1], Kartik K. Iyer[1], and Kalobaran Maiti[1*]

Date: 30 January 2024

Abstract

We study the single crystal growth and the electronic properties of TbIrIn$_5$. The single crystals of TbIrIn$_5$ were grown using the flux method. The analysis of the x-ray diffraction pattern shows high-quality single crystals formed in a tetragonal crystal structure. A trace of In-flux was also found in the data. Temperature and field-dependent magnetic measurements exhibit strong anisotropy and antiferromagnetic ordering with a Neel temperature of 43.2 K. The estimated Tb-moment from the data in the paramagnetic regime is found to be close to its free ion value. Interestingly, a strong diamagnetic behaviour and sharp resistivity drop typical of the superconducting phase is observed below 2.6 K which may have a link to the trace of indium present in the system. Further studies are required to ascertain this finding.

INTRODUCTIONS:

Heavy fermion intermetallic compounds like RTIn$_5$ (where R = Rare earth and T = transition metal) exhibit exceptional physical properties such as unconventional superconductivity, non-Fermi liquid behaviour, etc. arising from the interactions between the 4$f$ electrons and conduction electrons [1-4]. These materials crystallize ina tetragonal structure comprising of alternate stacks of RIn$_3$ and TIn$_2$ sub-lattices along the [001] direction leading to a pseudo-two-dimensional structure. While Co-based materials show superconductivity, Rh-based ones attracted attention due to interesting properties arising from the interplay between heavy fermion behaviour and antiferromagnetic ordering [5-8].Due to the large radial extension of the 5$d$ orbitals of In, the electron correlation of the $d$-electrons will be significantly weak in Ir compared to those in Co and Rh. On the other hand, the spin-orbit coupling will be stronger in this case. Thus, we investigated the growth and electronic properties of TbIrIn5 and found interesting results.

EXPERIMENTAL DETAILS:

Single crystals of TbIrIn$_5$ were grown by the self-flux method employing the flux-growth method. High-purity elements of Tb (4N), Ir (3N), and In (4N) were taken in a recrystallized alumina crucible in the ratio of 1:1:20, sealed in a quartz tube under vacuum, and heated at 1150ºC for 3 hours. The melt was then cooled to 600ºC at 10ºC per hour and centrifuged to eliminate In-flux.

The obtained single crystals were characterized for morphology and chemical compositions using an energy dispersive analysis of $x$-ray (EDX) spectrometer attached with a scanning electron microscope (SEM) (Zeiss ULTRAplus). The room temperature $x$-ray powder diffraction (XRD) study was carried out using a PANalytical $x$-ray diffractometer with Cu-K$_\alpha$ radiation, in the angle range of 10°<2θ<90°, and a step size of 0.02°. The Rietveld refinement of the experimental XRD data was done using the FullProf software package. The crystal orientation was determined by the Laue diffraction method.

Magnetization measurements were performed using a superconducting quantum interference device - vibrating sample magnetometer (SQUID-VSM) at different magnetic field intensities in the temperature range between 2 and 300 K. Temperature-dependent resistivity measurements were performed using PPMS (Quantum Design).

[1]Department of Condensed Matter Physics and Materials' Science, Tata Institute of Fundamental Research, Colaba, Mumbai 400005, India

*Corresponding author: kbmaiti@tifr.res.in

RESULT AND DISCUSSION:

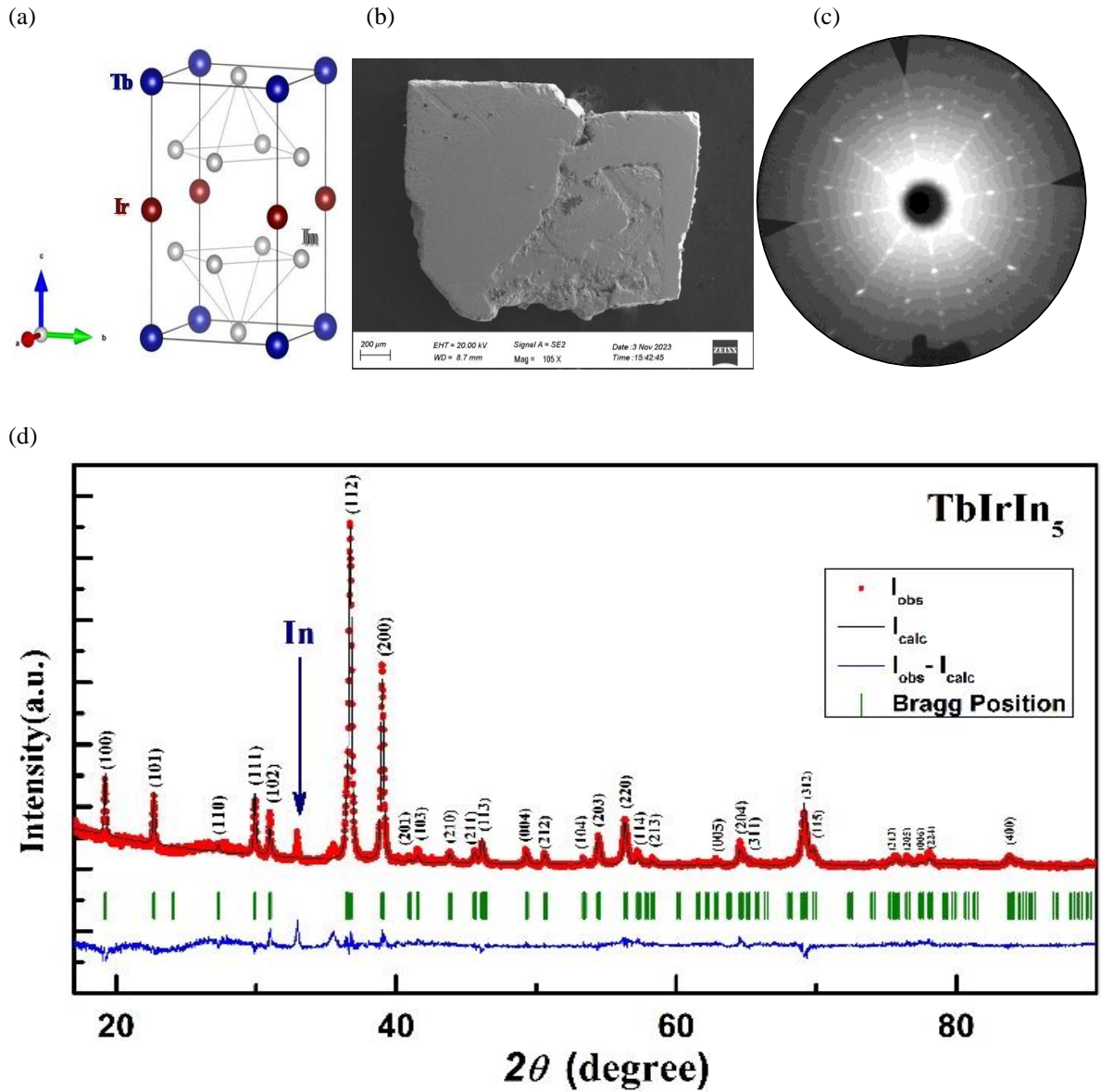

Fig. 1 (a) Crystal structure of TbIrIn$_5$. (b) SEM image of the sample exhibiting smooth flat surface of large size. (c) Laue diffraction pattern showing good crystallinity of the sample. (d) XRD pattern of the sample in powder form exhibiting single phase. Trace of a small amount of In-flux is also present in the sample.

The crystal structure of TbIrIn$_5$ is shown in Fig. 1(a) exhibiting a layered structure where Ir-layer is sandwiched by In-layers on both sides. While conduction layer in this effective two-dimensional structure is formed by the Ir-In layers, Tb also has strong hybridization with In which is important for the interesting physics of the material including possible heavy-fermion physics. SEM image of the sample is shown in Fig. 1(b) exhibiting nice flat regions of reasonably good size. The Laue diffraction pattern shown in Fig. 1(c) exhibit sharp well-ordered features suggesting excellent crystallinity of the sample. In order to study the structure further, we have powdered the sample and carried out XRD measurements. The experimental data is shown in Fig.1 (d) by red symbols. The Rietveld refinement of the experimental XRD data shown by line superimposed over the experimental data show a very good representation and suggests that TbIrIn$_5$ adopts the P4/mmm tetragonal HoCoGa$_5$-type structure. The structure parameters obtained from the refinement are given in the Table 1.

Table 1. Structural parameters of TbRhIn$_5$

| Crystal Structure | Tetragonal | Atomic Positions | x | Y | z |
|---|---|---|---|---|---|
| a (Å) | 4.6168(3) | Tb | 1a | 0 | 0 | 0 |
| c (Å) | 7.3923(4) | Ir | 1b | 0 | 0 | 0.5 |
| α = β = γ | 90° | In1 | 4i | 0.5 | 0.5 | 0 |
| Space group | P4/mmm | In2 | 1c | 0 | 0.5 | 0.3020(3) |

In Fig. 2(a), we show the temperature-dependent magnetic susceptibility, χ measured with the external applied field along the [100] and [001] directions. The experimental data exhibit large anisotropy with the moments significantly larger for the magnetization along [001] direction than [100] direction. This suggests [001] as the easy axis of magnetization. A sharp antiferromagnetic (AFM) transition is observed at a Néel temperature, T$_N$ of 43.2K; the transition is sharp for the field along [001] direction. For the measurements along [100] direction, the AFM transition is observed at the same temperature. However, instead of a reduction of the moment at lower temperatures due to AFM order, a significant increase in χ is observed at lower temperatures. Such increase may be attributed to the disorder and/or imperfections present in the system as also observed in other materials [9,10]. The χ$^{-1}$ vs temperature data, shown in Fig. 2(b), exhibits a linear temperature dependence above T$_N$, typical of the Curie-Weiss behaviour for a large temperature range. Estimated effective magnetic moment, $\mu_{eff} = 10.01\ \mu_B/$Tb which is close to the free ion value of 9.72 $\mu_B$/Tb. The negative $\theta_P\ (= -15\ K)$ suggests antiferromagnetic interaction among the moments, which is consistent with the observation of AFM ordering at lower temperatures. For T<T$_N$, χ along [001] decreases significantly with cooling. But in the

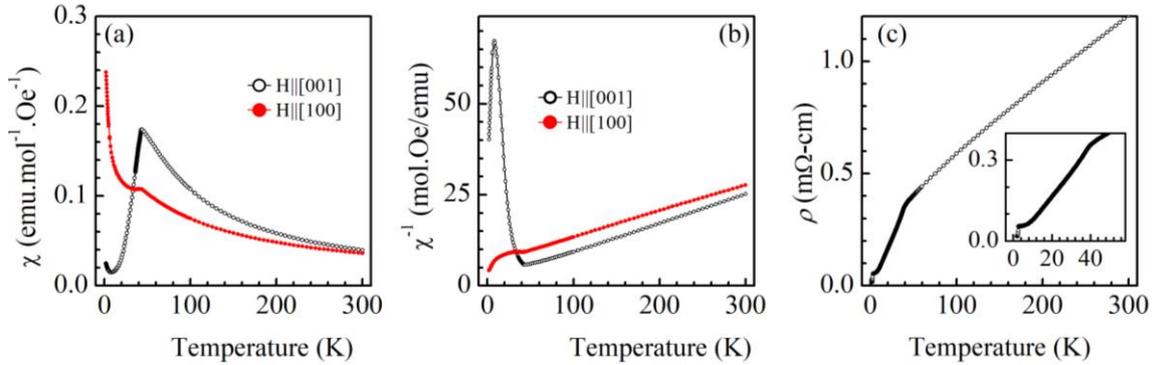

Fig.2 (a) Magnetic susceptibility, χ vs Temperature data for the magnetic field, H along [100] and [001] directions. (b) Inverse magnetic susceptibility. (c) Temperature dependence of the in-plane electrical resistivity, ρ. The inset in (c) shows the expanded part of the data at low temperatures.

[100] direction, χ increases with cooling which is unusual for the AFM ground state. At T < 2.6K, χ decreases sharply and the moment becomes negative indicating the signature of superconductivity in this material.

In Fig 2(c), we show the temperature dependence of the *ab*-plane electrical resistivity. The data show close to linear decrease in resistivity with temperature which is different from the expected behaviour of a Fermi liquid. A sharp decrease is observed below *T$_N$* due to magnetic long-range order as also observed in various other rare-earth intermetallics [11]. Within the magnetically ordered state, the resistivity shows again a nearly linear behaviour down to about 10 K and then flattens up at lower temperatures. Interestingly, a superconducting transition is observed at about 2.6 K consistent with the magnetization data. While the discovery of superconductivity in such highly correlated strong spin-orbit coupled system could open up a new paradigm in the study of superconductivity, in particular, in the heavy fermion class of systems, the superconducting phase observed here may also be related to the trace of indium flux present in the sample though the transition temperature of bulk indium is slightly higher (= 3.4 K) [12]. Further studies are required to ascertain this phenomenon.

CONCLUSIONS:

We have grown high-quality single crystals of TbIrIn$_5$. Magnetic properties show strong anisotropy with an easy axis along [001] direction and sharp anti-ferromagnetic transition at 42.5 K. The anisotropy at low temperatures is significantly high with a reversed trend of larger in-plane moment than that along [001]-direction. The estimated moment from the paramagnetic data is close to the free ion value of Tb. Interestingly, we observe a signature of superconducting transition at 2.6 K. Further studies are required to ascertain this finding as such behavior may also arise from the trace of In-flux remaining in the sample as impurities.

The authors acknowledge financial support from DAE, Government of India (Project Identification No. RTI4003, DAE OM No. 1303/2/2019/R&D-II/DAE/2079).